\documentclass[12pt,a4paper]{panl}
\usepackage{cite}
\usepackage{wrapfig}
\usepackage{graphicx}
\usepackage{amssymb}
\usepackage{amsfonts}
\usepackage{amsmath}
\usepackage{longtable}
\usepackage{rotating}
\usepackage{lscape}
\usepackage{epsfig}
\usepackage{multirow} 
\newcommand{\lr}[1]{ \left( #1 \right) }
\newcommand{\lrs}[1]{ \left[ #1 \right] }
\newcommand{\lrc}[1]{ \left\{ #1 \right\} }

\newcommand{\Tr}{{\mathrm{Tr}\,}}
\newcommand{\ReP}{{\mathrm{Re}\,}}

\originalTeX
\begin{document}

\title{Studying properties of the SU(2) QCD by lattice field theory methods}
\maketitle
\authors{I.\,Kudrov$^{a}$\footnote{E-mail: ilya.kudrov@ihep.ru},
V.\,Bornyakov$^{a,b}$,
V.\,Goy$^{c, d}$}
\setcounter{footnote}{0}
\from{$^{a}$\, NRC “Kurchatov Institute”- IHEP, Protvino, 142281 Russia}
\from{$^{b}$\, NRC ”Kurchatov Institute”,  Moscow, 123182 Russia}
\from{$^{c}$\, Pacific Quantum Center, Far Eastern Federal University, 690950 Vladivostok,  Russia}
\from{$^{d}$\, Institute of Automation and Control Processes, Far Eastern Branch, Russian Academy of Science, 5 Radio Str., 690041 Vladivostok, Russia}

\begin{abstract}

We present new results on properties of $SU(2)$ QCD in lattice regularization. 
Our main goal is to find the transition line confinement - deconfinement in $\mu - T$
plane. We compute the Polyakov loop and the string tension to determine this line.
\end{abstract}
\vspace*{6pt}


\label{sec:intro}
\section*{Introduction}

Many  physical phenomena, e.g. heavy ion collisions \cite{Andronic:2017pug}, hot matter in the early universe, or neutron stars require an understanding of the QCD phase diagram. The region of the phase diagram along the temperature axis is extensively studied using first principle lattice QCD calculations as well as perturbative calculations at very high temperatures. However, the addition of a baryon chemical potential introduces the so-called sign problem for lattice QCD methods \cite{Nagata:2021ugx}. The lattice QCD methods have been developed, which allow studying the phase diagram in the vicinity of a zero baryon chemical potential \cite{Ratti:2019eY}, but for large chemical potential, those methods produce uncontrollable errors. For large baryon chemical potential, it is possible to use effective models and other approaches, but they make some assumptions which lead to uncontrollable errors and therefore have limited areas of applicability. Such complications do not arise if one considers QCD with two colors \cite{Kogut:2000ek, Braguta:2023yhd,Itou:2025vcy}.

$SU(2)$ QCD phase diagram has both similarities and differences with that in $SU(3)$  case. In $SU(2)$  QCD baryons are bosons and consist of two quarks instead of three. It leads to a different particle spectrum \cite{Hands:1999md} and phenomena like the appearance of a diquark condensate at high baryon density \cite{Cotter:2012mb,Astrakhantsev:2020tdl,Iida:2024irv }. The important  similarity is the presence of a confinement-deconfinement transition. For zero chemical potential, it is known that the transition is a crossover \cite{Cotter:2012mb}, and at low temperature and large chemical potential the transition was studied in \cite{Bornyakov:2018sab, Begun:2022bxj}.

In this work we focus on the confinement-deconfinement transition line at temperature $T \ge 170$ MeV and non-zero quark chemical potential. Already existing experimental facilities, such as LHC, RHIC and NICA, will be used to intensively probe regions of the QCD phase diagram, where confinement-deconfinement transition might be observed. Since both $SU(2)$  and $SU(3)$  theories exhibit confinement-deconfinement transition, $SU(2)$ QCD might be useful for testing methods of finding this line and then apply these methods to the $SU(3)$  case.

Despite the fact that in $SU(2)$  QCD the confinement-deconfinement  transition  was studied at non-zero  chemical potential before \cite{Cotter:2012mb,Boz:2019enj,Bornyakov:2018sab,Begun:2022bxj}, some of its features still remain unclear. Namely, existence of the confinement-deconfinement transition at very low temperature and large chemical potential, the exact path of the transition line, and the order of the transition. The location of the confinement-deconfinement transition was found in Ref.~\cite{Begun:2022bxj} for temperatures $ 100 <  T < 140$ Mev. In this work we want to extend this line to higher $T$/lower $\mu$.   

\section{Simulation details}
\label{sec:sumulation} 

We  use lattices with fixed spatial size $N_s=32$ and vary the temporal size $N_t$ from 10 to 24 at nonzero quark chemical potential $\mu$ and from 4 to 26 at $\mu=0$.
The chemical potential $\mu$ varies in the range  $ 0 \le a\mu \le 0.2$.
The lattice gauge field configurations were generated with the tree level  improved Symanzik gauge action \cite{Weisz:1982zw} and improved staggered fermion action with a diquark source term \cite{Hands:1999md}:
\begin{equation}
  S_{SU(2)QCD}=S_G+S_{stag}\,,
\end{equation}
where 
\begin{equation}
  S_G  =  \frac{\beta}{2} \lr{c_0 \underset{plaq}{\sum}\, \ReP\Tr\lr{1-U_{plaq}} + c_1 \underset{rt}{\sum}\, \ReP\Tr\lr{1-U_{rt}}}\,, 
\end{equation}
\begin{equation}
\label{eq:fermionic_action}
\nonumber
  S_{stag}  =  \underset{x}{\sum} \bar{\psi}_{x}
    \lrs{\underset{\nu}{\sum}\frac{\eta_{x,\nu}}{2}\lrc{
    \tilde{U}_{x,\nu} e^{\delta_{\nu,0}\mu a} \psi_{x+\hat\nu} -
    \tilde{U}_{x-\hat\nu,\nu}^{\dagger}  e^{-\delta_{\nu,0}\mu a} \psi_{x-\hat\nu}}
    + m_q a\; \psi_{x}}
    \nonumber
    \end{equation}
    \begin{equation}
  \hspace{-3.0cm} +  \underset{x}{\sum}\frac{1}{2}\lambda \lrs{\psi_x^T \sigma_2 \psi_x + \bar{\psi}_x \sigma_2 \bar{\psi}_x^T}\,,
\end{equation}
where $c_0$, $c_1$ -- parameters of improved lattice gauge action, $\beta$ -- inverse coupling constant, $U_{x,\mu}$ -- $SU(2)$ link variable, $\tilde{U}_{x,\mu}$ - stout smeared link variable \cite{Morningstar:2003gk}, $\eta_{x,\mu}$ -- staggered sign function. $S_{stag}$ has implicit summation over the flavor index.

The lattice action parameters are the same as in Ref.~\cite{Begun:2022bxj}: $\beta=1.75$, quark mass in lattice units $am_q=0.0075$, the diquark source 
term coupling $\lambda=0.00075$. In  Ref.~\cite{Begun:2022bxj} it was found that $a = 0.048(1)$~fm and the pion mass is 680(40) MeV for these parameters if the physical scale is fixed by Sommer parameter $r_0 = 0.468(4)$~fm. We used about 4,000 configurations at every $(T,\mu)$ point. The bootstrap method was employed  to compute the statistical errors.

\section{Polyakov loop}
\label{sec:PL}
The Polyakov loop defined as
\begin{equation}
    P(\vec{x}) = \frac{1}{2} Tr \prod_{x_4=1}^{N_t} U_4(\vec{x},x_4)
\end{equation}
is an order parameter for the confinement-deconfinement transition in gluodynamics.
The Polyakov loop susceptibility $\chi_P$ 
\begin{equation}
 \chi_P = <P^2> - <P>^2 
\end{equation}
is used to find the transition temperature.
In the theory with quarks, it is still used to locate the transition. For this purpose
inflection point or the static source entropy $S_q$ 
\begin{equation}
S_q = -\frac{\partial F_q}{\partial T} 
\end{equation}
where $F_q = -T \log(<P>)$ - static quark free energy, are also used. 

The Polyakov loop needs to be renormalized. We used a two steps procedure. First, we applied 10 sweeps of the hypercubic blocking  \cite{Hasenfratz:2001hp} to the links $U_4(x)$ which reduces the divergent self-energy of the static quark. Then we define
the renormalization factor $z$ as
\begin{equation}
<P>_{ren}(N_t=4) =  1 = z^4 <P>(N_t=4),
\end{equation}
where $<P>(N_t=4) $ is the Polyakov loop at the highest temperature and zero chemical potential. 
We find $z=1.017$. 
It is assumed that $z$ is independent of $\mu$, thus the renormalized $ <P>_{ren}(N_t)$
is defined as
\begin{equation}
<P>_{ren}(N_t) =  z^{N_t} <P>(N_t)
\end{equation}
for any $\mu$.
We use $<P>_{ren}$ to compute the  susceptibility $\chi_P$.
In Fig.~\ref{fig:susceptibility}(left) we present our results for $\chi_P$ as a function of temperature for different $\mu$. We used a Gaussian fit to determine the position of the maximum. One can see that the height of the maximum becomes lower with increasing $\mu$. This suggests that the transition becomes weaker for higher $\mu$.  
\begin{figure}[t]
\vspace{-0.6cm}
\begin{center}
\hspace{-0.9cm}
\includegraphics[width=76mm]{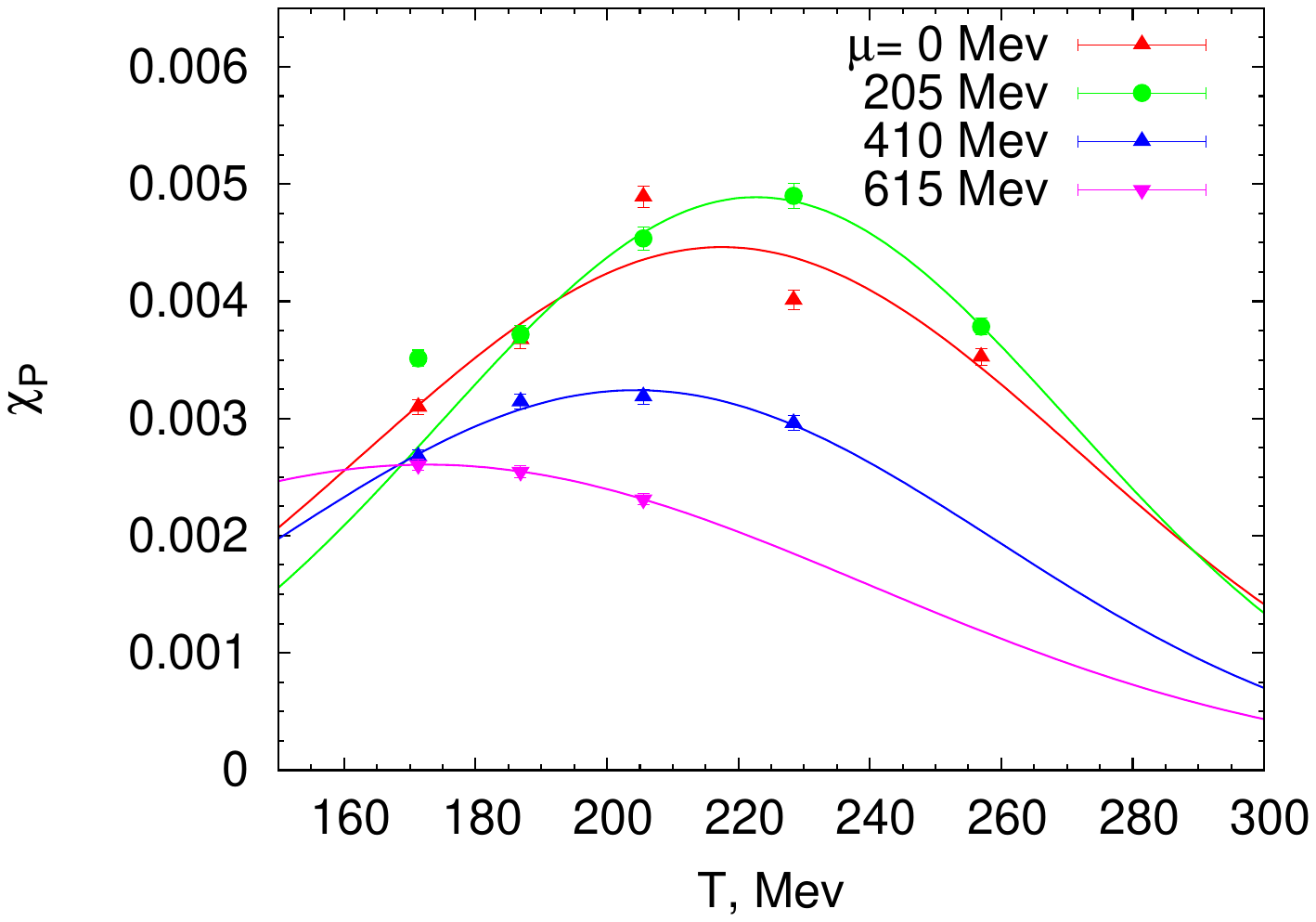}
\hspace{-1.6cm}
\includegraphics[width=76mm]{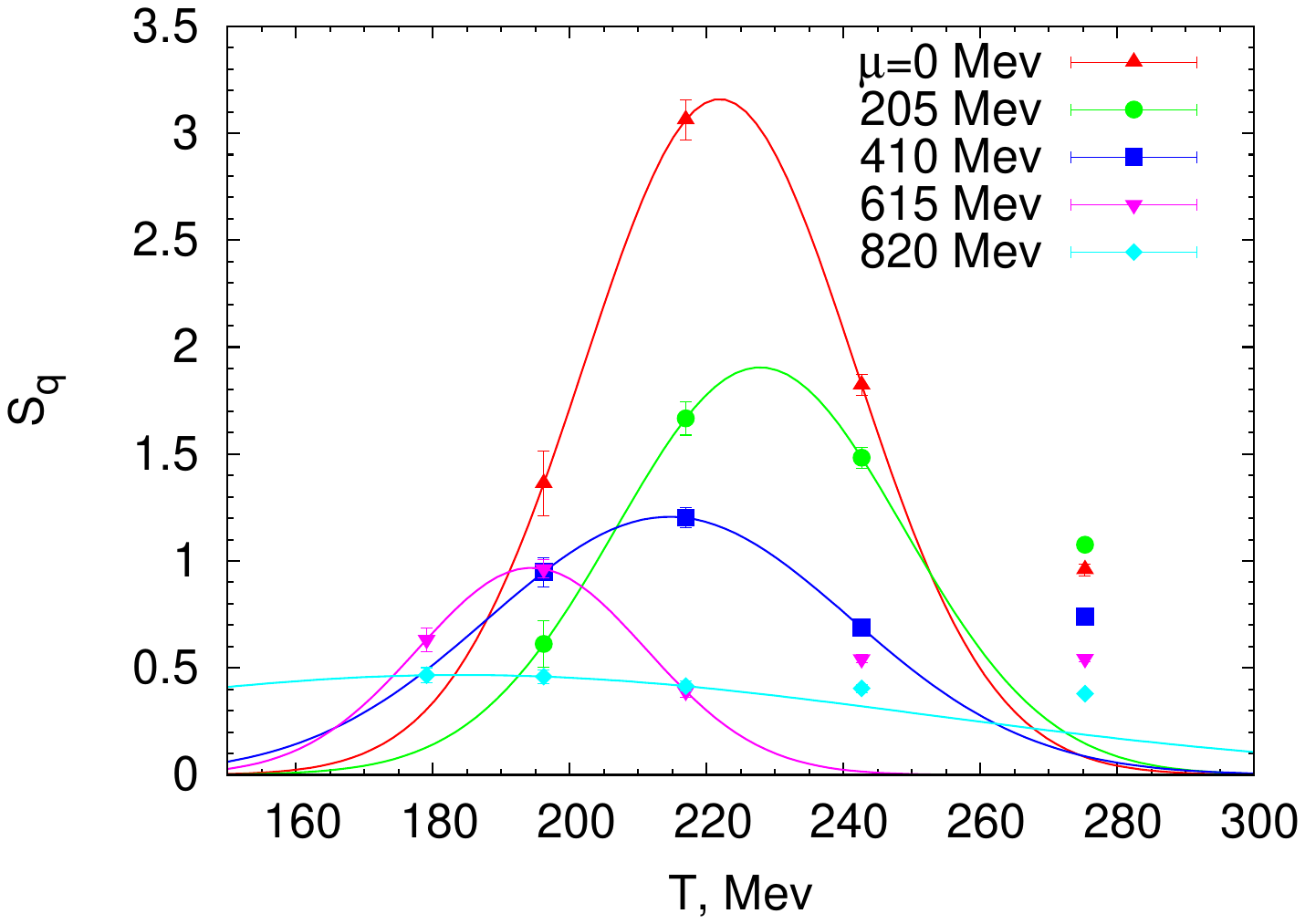}
\vspace{-3mm}
\caption{Left:The Polyakov loop susceptibility vs. $T$; Right: The static quark entropy vs. $T$. The curves show fits to gaussian function.}

\end{center}
\labelf{fig:susceptibility}
\vspace{-5mm}
\end{figure}
In Fig.~\ref{fig:susceptibility}(right) respective results for the static quark entropy $S_q$ are shown. Again, we use a Gaussian fit to determine the position of the maximum. In agreement with data for the susceptibility, we see that the transition becomes weaker for higher $\mu$.

\section{String tension}
\label{sec:ST}
The string tension $\sigma$ is an obvious observable indicating if the hadron matter under study is in the confinement or deconfinement phase.  In QCD at $T=0$ due to the presence of sea quarks
the string breaks down at a distance $\approx$  1.25 fm \cite{Bali:2005fu}, where
the total energy of the QCD string becomes large enough to 
create a quark-antiquark pair from the vacuum. Still, one can measure the string tension using the static potential extracted from the Wilson loops due to small overlap with the broken string (i.e. two static-light mesons) state.  At nonzero temperature the string tension $\sigma(T)$ was measured in the gluodynamics in Refs.~\cite{Kaczmarek:1999mm,Cardoso:2011hh} and in QCD in Ref.~\cite{DIK:2004xwj}. In all cases the Polyakov loop correlator was used to compute $\sigma(T)$. In the work on $SU(2)$ QCD  \cite{Begun:2022bxj} the Wilson loop was used to extract the static potential and to estimate the respective string tension to determine the confinement-deconfinement transition at low temperatures, $ 100 < T <  140$ MeV.  Here we extend this method, with some modification, onto higher temperatures.

We compute the Wilson loops $W(r,t)$ for ranges $1 \le r/a \le N_s/2$, $1 \le t/a \le N_t$ using the hypercubic blocking for the temporal links $U_4(x)$ and APE smearing for
the spatial links $U_i(x)$ \cite{Hasenfratz:2001hp}. The Wilson loops at finite temperature were studied recently in Refs.~\cite{Bala:2019cqu,Bala:2021fkm,Ali:2025iux}. It was shown that at finite temperature, the sharp  delta peak of the spectral function acquires a thermal width, making the extraction of the thermal static 
 potential from the Wilson loop much more difficult in comparison with $T=0$ case. The nonzero width implies a nonzero imaginary part of the thermal potential. The proper fit functions for $W(r,t)$ were suggested which allow one to extract the real and imaginary parts of the thermal potential. We do not use them here since we were not able to obtain stable fits. We will do this work in the future after improving our statistics. Here we use a simple expression 
\begin{equation}
W(r,t) = A (e^{-V(r)t}+ Be^{-E_1(r)t}) 
\end{equation}
and make a fit of the effective potential 
\begin{equation}
    aV_t(r) = \log (W(r,t)/W(r,t+a))
    \label{eq:Vt}
\end{equation} 
for small $4<t/a<6$ with three fit parameters $aV(r), aE_1(r), B$. 

\begin{figure}[t]
\vspace{-6mm}
\begin{center}
\hspace{-0.9cm}
\includegraphics[width=76mm]{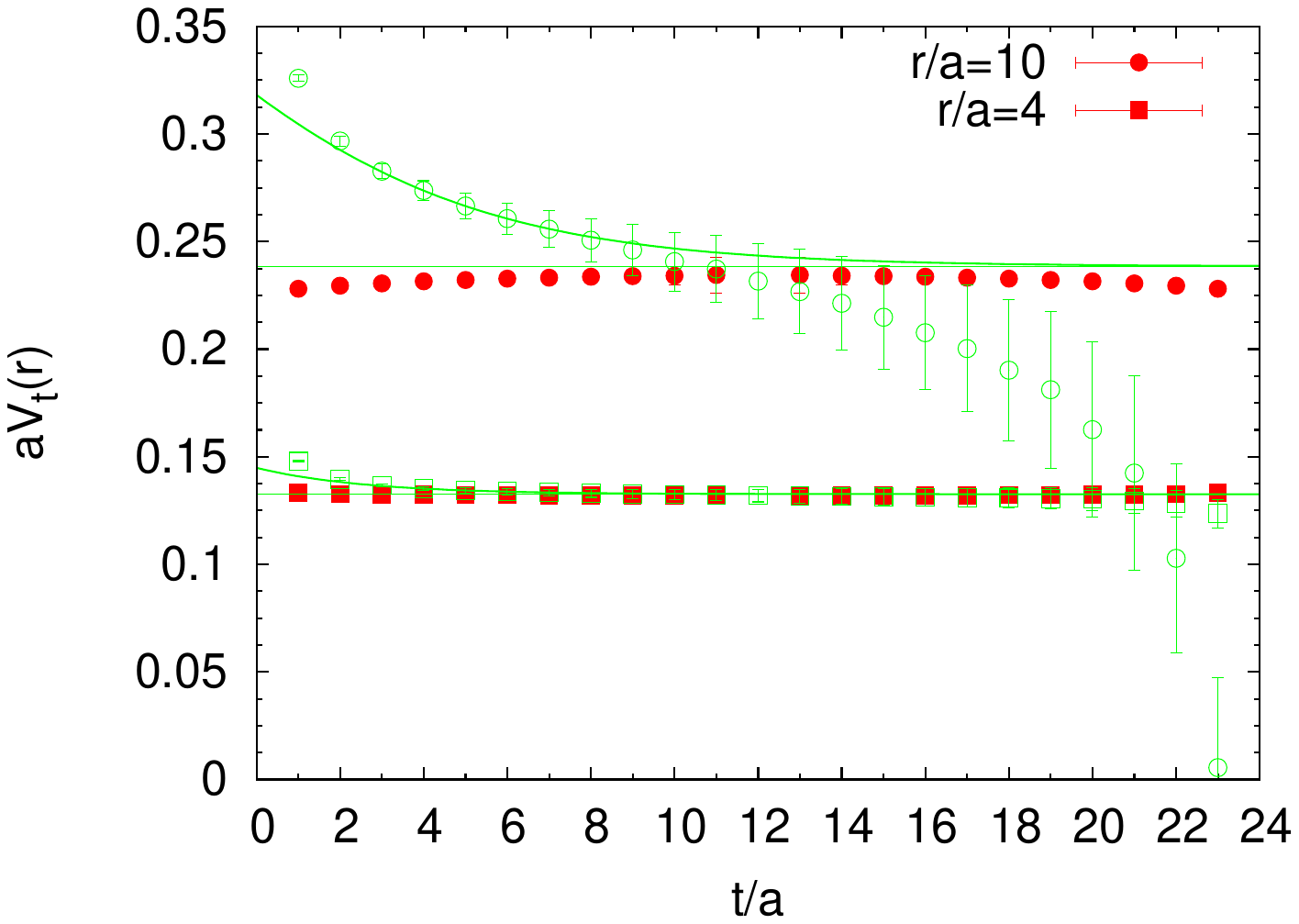}
\hspace{-1.6cm}
\includegraphics[width=76mm]{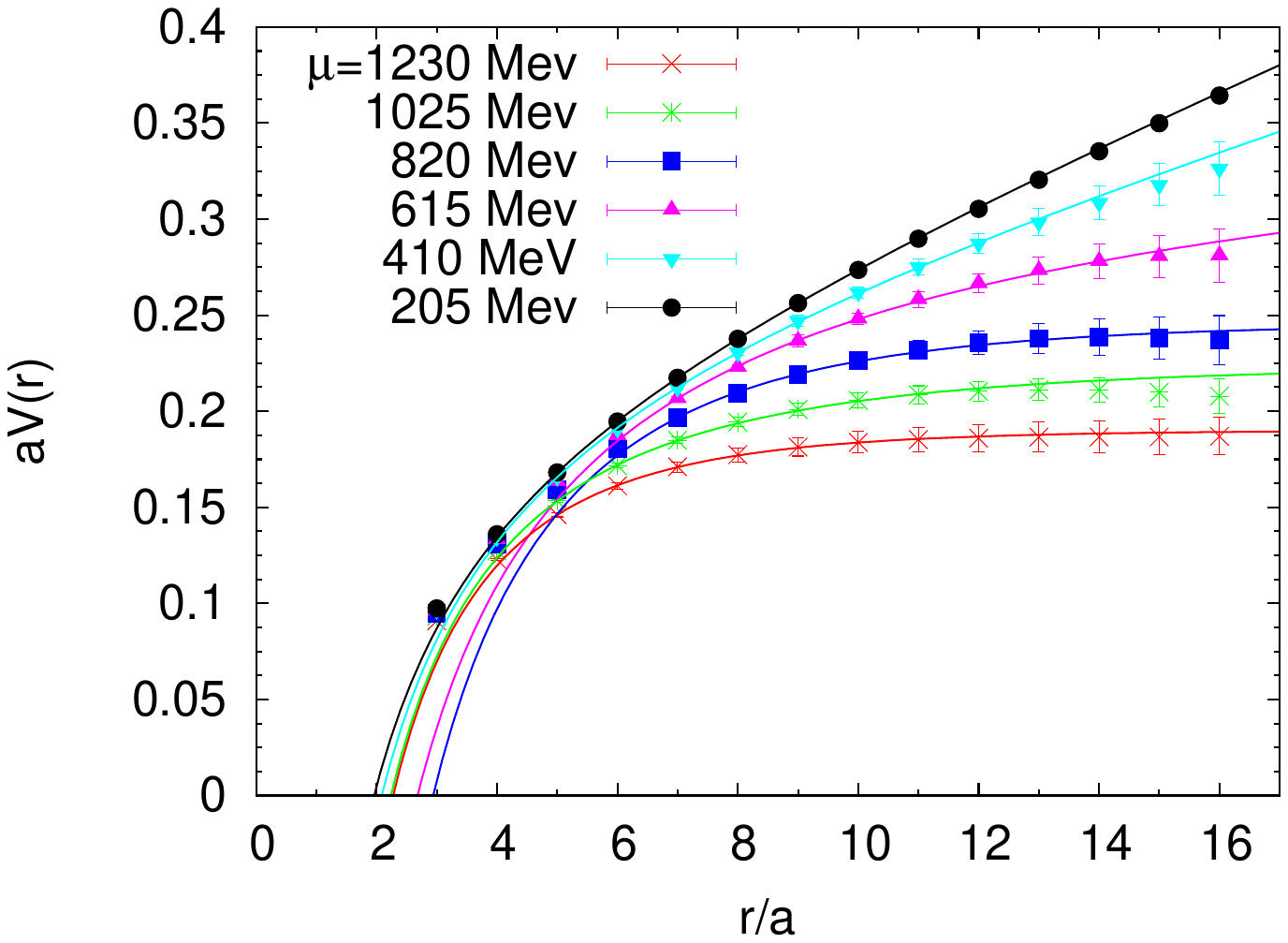}
\vspace{-3mm}
\caption{Left: The thermal potential $V_t(r)$ vs. $t$ for $r/a=4$ and 10 computed via eq.~(\ref{eq:Vt}) (empty symbols)  and via  eq.~(\ref{eq:bala}) (filled symbols) at $T=171$ MeV, $\mu=615$ MeV; Right: The thermal potential $V(r)$ at $T=171$~MeV for few values of $\mu$. The lines are fits to eq.~(\ref{eq:Cornell}) (for low $\mu$) or fits to  eq.~(\ref{eq:screened}) (for high $\mu$).}
\end{center}
\labelf{fig:potential}
\vspace{-5mm}
\end{figure}

In Fig.~\ref{fig:potential}(left) we show $aV_t(r) $ 
 as a function of $t/a$ for two values of $r/a$ computed at $T=171$ MeV, 
 $\mu=615$ MeV. 
 In Ref.~\cite{Bala:2019cqu} the following empirical form of  $V_t(r) $
 was suggested
 \begin{equation}
    a\tilde{V}_t(r) = \frac{1}{2} \log (W(r,t)/W(r,aN_t-t))/(N_t/2-t/a)
    \label{eq:bala}
\end{equation} 
We compare $\tilde{V}_t(r)$ with $V_t(r) $ in Fig.~\ref{fig:potential}. One can see that parameter $aV(r)$  of the fit (\ref{eq:Vt}) is in agreement with the prediction
of eq.~(\ref{eq:bala}) computed at $t/a=N_t/2-1$. We find this agreement for other values of $T$ and $\mu$ as well.

In Fig.~\ref{fig:potential}(right) the potential $aV(r)$ for  $T=171$ MeV and a few values of $\mu$ is depicted. The fits shown in the figure are to the Cornell potential: 
\begin{equation}
    V(r) = V_0+\sigma r + \alpha/r \,,
\label{eq:Cornell}
\end{equation}
or to a screened potential
\begin{equation}
    V(r) = V_0+\frac{\alpha}{r} e^{-m_{scr}r} \,.
\label{eq:screened}
\end{equation}

One can see that the slope of the thermal potential $V(r)$ decreases with increasing $T$ or $\mu$. $\sigma(T)$ for various values of $\mu$ is presented in Fig.~\ref{fig:ph_diag}(left). We use extrapolation to estimate $\mu$ at which the string tension approaches zero.  

All results for confinement-deconfinement transition temperature $T_d$ obtained with the inflection point, Polyakov loop susceptibility, static source entropy and string tension  are shown in Fig.~\ref{fig:ph_diag}(right). We took a fairly conservative error estimate  for these results: half the distance between our data points included in the corresponding fits. The results for all four observables show qualitatively similar behavior. The difference between values of $T_{d}(\mu)$ obtained for different observables is within the error bars. The data show slow decreasing of  $T_{d}$ at small $\mu$ and roughly linear decreasing at $\mu > 200$ MeV. In Fig.~\ref{fig:ph_diag}(right) we also show $T_d$ values obtained in Ref.~\cite{Begun:2022bxj} for the lower temperatures. It is clear that new results for $T_d$ are in qualitative agreement with results of Ref.~\cite{Begun:2022bxj}.
 Our value for $T_{d}$ at $\mu=0$ is in good agreement with the values obtained earlier in Refs.~\cite{Cotter:2012mb,Boz:2019enj}. There is also agreement with results of Ref.~\cite{Boz:2019enj} at $\mu  < 500$ MeV. But at higher $\mu$ results of  Ref.~\cite{Boz:2019enj} show independence of $T_{d}$ on $\mu$ contrary to our results.
\begin{figure}[t]
\vspace{-6mm}
\begin{center}
\hspace{-0.9cm}
\includegraphics[width=76mm]{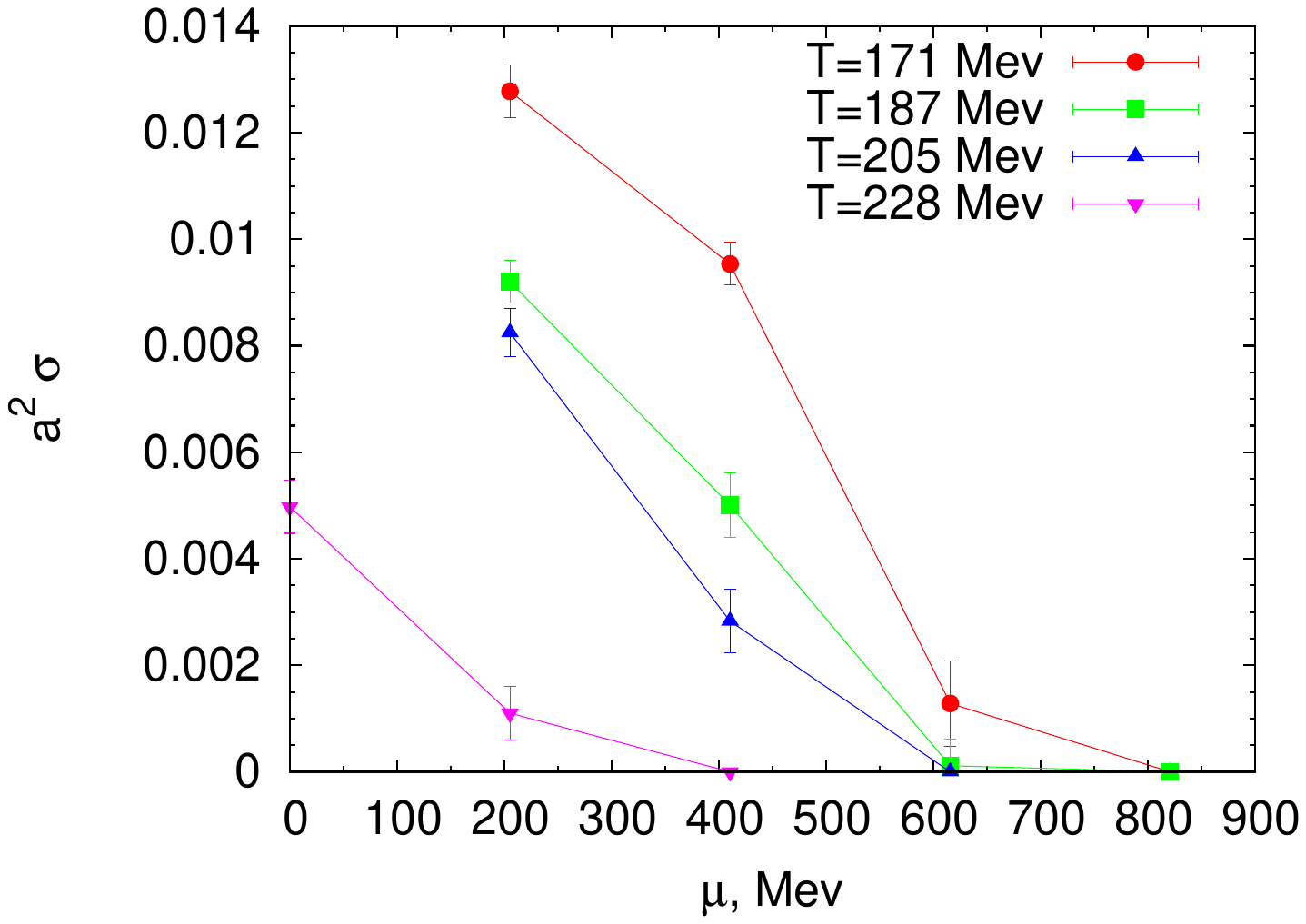}
\hspace{-1.6cm}
\includegraphics[width=76mm]{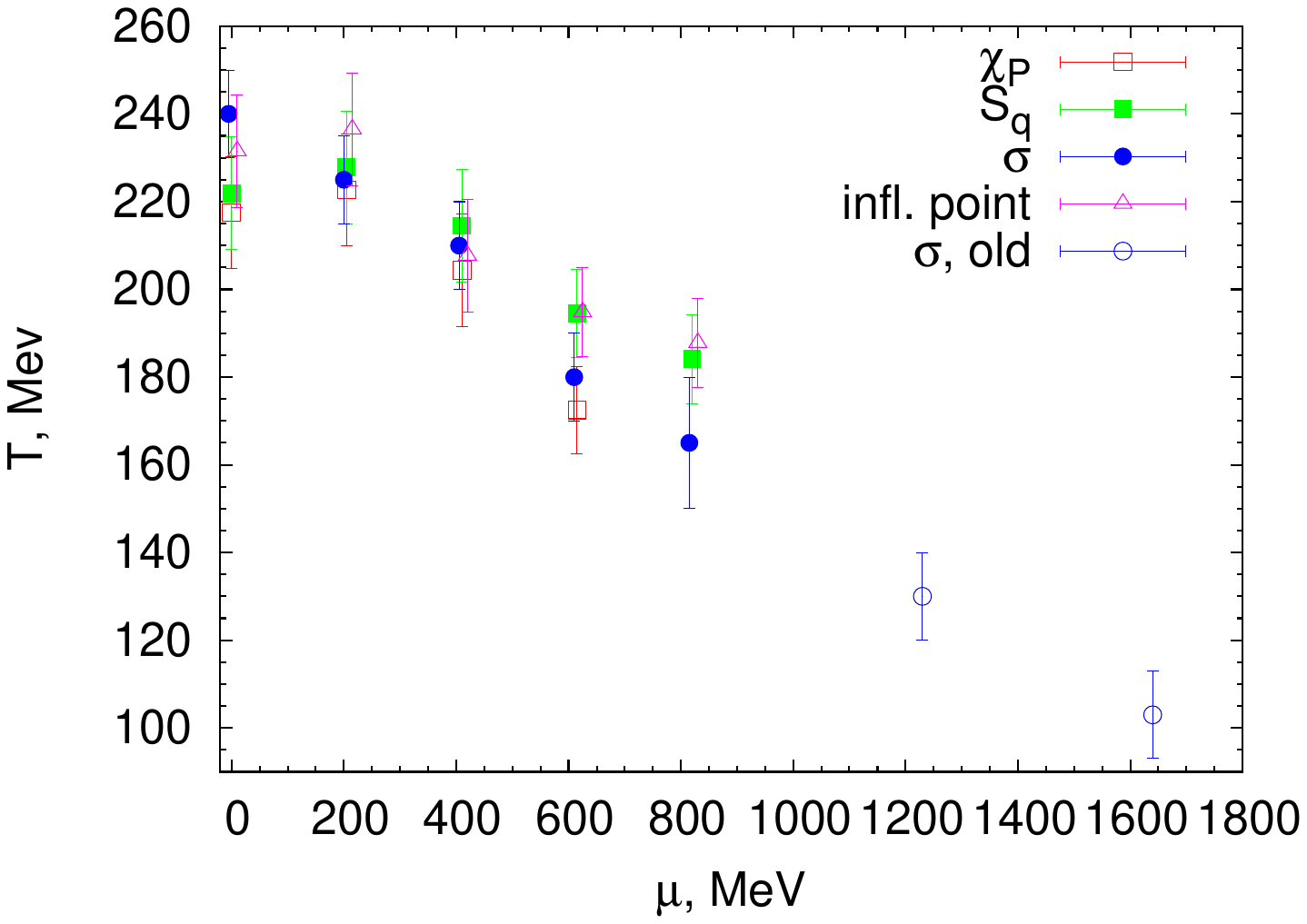}
\vspace{-3mm}
\caption{Left: The string tension vs. $\mu$ for various $T$; Right: phase diagram}
\end{center}
\labelf{fig:ph_diag}
\vspace{-5mm}
\end{figure}

\section{Conclusions}
\label{sec:conclusions}
In this work we determined  the position of the confinement-deconfinement transition line  in the plane $\mu - T$ in SU(2) QCD for the chemical potential range $0 - 830$ MeV.
We used the Polyakov loop inflection point, its susceptibility $\chi_P$ and the static quark entropy $S_q$ to determine the transition line. Additionally, we computed the string tension $\sigma(T)$ using the Wilson loop to compute the thermal static potential. These four observables produced results for the transition temperature $T_d(\mu)$ compatible within error bars.  Our estimate for $T_d$ at $\mu=0$ is $T_d(0) = 230(10)$ MeV or, equivalently, $r_0 T_d(0) = 0.55(4)$. Our results indicate that $T_d(\mu)$ changes slowly at $\mu < 200$ MeV and then it changes rather fast with almost 
linear dependence on $\mu$. Our results are in qualitative agreement with results of Ref.~\cite{Begun:2022bxj} where the same lattice action with the same set of parameters was used to compute $T_d$ at higher $\mu$. At the same time, we note a disagreement with values for  $T_d(\mu)$ obtained in Ref.~\cite{Boz:2019enj} in the range  $200 < \mu < 800 $ MeV.

It is worth noting that our results for $\chi_P$ and $S_q$ indicate that the confinement-deconfinement transition, which is a crossover, becomes weaker with increasing $\mu$. This is an important observation in view of the still unclear fate of this transition in the limit $T \to 0$. One possible scenario is that $T_d$ becomes independent of $\mu$ at even higher $\mu$. This deserves further study.


\label{sec:acknowledgement}
\section*{Acknowledgement}
Computer simulations were performed on the computing cluster of Far Eastern Federal University ,
the Central Linux Cluster of the NRC “Kurchatov Institute”-IHEP, and the Linux Cluster of  KCTEP NRC “Kurchatov Institute”. Resources of the federal collective usage center Complex for Simulation and Data Processing for Mega-Science Facilities at NRC ”Kurchatov Institute” were also used.

\label{sec:funding}
\section*{Funding}
This research was funded by the Russian Science Foundation (Grant 23-12-00072). 
The generation of gauge field configurations on lattices with $N_t=12, 14, 16$ and 18 (using the computing cluster of Far Eastern Federal University)  completed by V.G. was supported by Grant No. FZNS-2024-0002 of the Ministry of Science and Higher Education of Russia.

\bibliographystyle{pepan}
\bibliography{pepan_biblio}

\end{document}